# Different Neurons Population Distribution correlates with Topologic-Temporal Dynamic Acoustic Information Flow

**Walter Riofrio[1] and Luis Angel Aguilar[1, 2]**

[1] Neuroscience and Behaviour Division,
Universidad Peruana Cayetano Heredia,
Walter.Riofrio@terra.com.pe

[2] Castilla y Leon Neuroscience Institute,
Salamanca, Spain
LUCHO@usal.es

In this study, we will focus on two aspects of neural interconnections. One is the way in which the information flow is produced, and the other has to do with the neural distribution with specific architectural arrangements in the brain. It is very important to realize that both aspects are related, but it is possible to support in the former that the information flow is not *only* governed by the number of spikes in the neurons, but by a series of other factors as well. Here we show the role played by GABAergic neurons in acoustic information transmission in the Central Nucleus of Inferior Colliculus (CNIC). The discovery and distribution of GABAergic neurons in the CNIC improve our understanding on how the inhibitory actions of neurotransmitters participate in the information flow. We report a neural spatial-temporal cluster distribution, associated with each isofrequency region. With these results, we will shed some light onto the emergence of certain mental properties starting from the neural dynamic interactions.

## 1. Introduction

The Inferior Colliculus (IC) is the processing center of ascending information to the thalamus and brain cortex. It controls the nucleus of the lower auditory pathway, and it plays an important role in multi-sensorial integration and in motor-auditory reflex production [19, 23]. These different functions must be a consequence of the integration of excitatory and inhibitory signals from the ascendant, descendant, and commissural projections [11, 18, 24].

The inhibitory action of GABA and Gly is what shapes the temporal and spectral response of IC neurons. Due to the relevance of inhibitory processes in the physiologic responses of the IC neurons, it is realistic to think that we might obtain some explanations of the acoustic information flow dynamics from the neural populations and the histochemical and cytoarchitectural organization of this nucleus.

Tonotopic organization is a fundamental property of the auditory system [26], and it is now well established that the morphological substrate for such tonotopy in the CNIC is its laminar organization [20].

It is further known that the *rate code* measures the intensity of a signal by means of the number of spikes on a single neuron or a population of neurons over a period of time [2]. Once the rate code has been determined, we take into account the total number of spikes over a period of time, but the order or timing of the spikes is not relevant. For a considerable period of time, it was thought that the variability in the inter-spike intervals would limit the sensory information that neurons would transmit.

Notwithstanding, recent studies indicate that variability would be involved in certain important advantages in information flow: variability, or noise, could enhance sensitivity to weak signals, a phenomenon called *stochastic resonance* [6].

Others suggest additional important advantages of variability or noise. They put forward that this apparent spike generation variability by neurons represents signals from excitatory postsynaptic potential (EPSPs), which is commonly referred to as a *temporal code* [12]. The main idea of using the temporal code concept is that the exact timing of nerve spikes carries with it more information than the rate alone.

This research into information flow tells us that the so-called 'noise' associated with variability in neural inter-spikes would have correlations with cognitive processes. Even though researchers point out the need for more inquiry in this direction, a common agreement has been produced recently: at the moment, the temporal notion is a very important component for understanding the way in which information flow is transmitted, and comprehending this phenomenon will have strong implications on the studies of mental processes [1, 3, 25].

In this paper, we propose a working hypothesis on the implications of the sensorial information flow through the neural network and its neural distribution with regard to the most basic capacities involved in mental or cognitive properties.

## 2. Cytoarchitectural Organization in Central Nucleus of Inferior Culliculus

In a recent study [15] about the distribution of GABA-IR and GABA-IN neurons across and within the frequency-band laminae, we found that there are neurochemical differences with regard to the tonotopic organization of the CNIC (see Methods in reference 15): "…differences within the laminae are greater along the dorsomedial–ventrolateral axis than along the rostro-caudal axis…differences across frequency regions are minor…" [15, page 920].

Although, we studied the quantitative anatomical distribution of GABAergic neurons in each portion of the IC, the CNIC in rats possesses up to 30% of GABAergic neurons [15, see page 921].

The presence of GABA-IR and GABA-IN throughout the entire IC is shown in fig. 1.

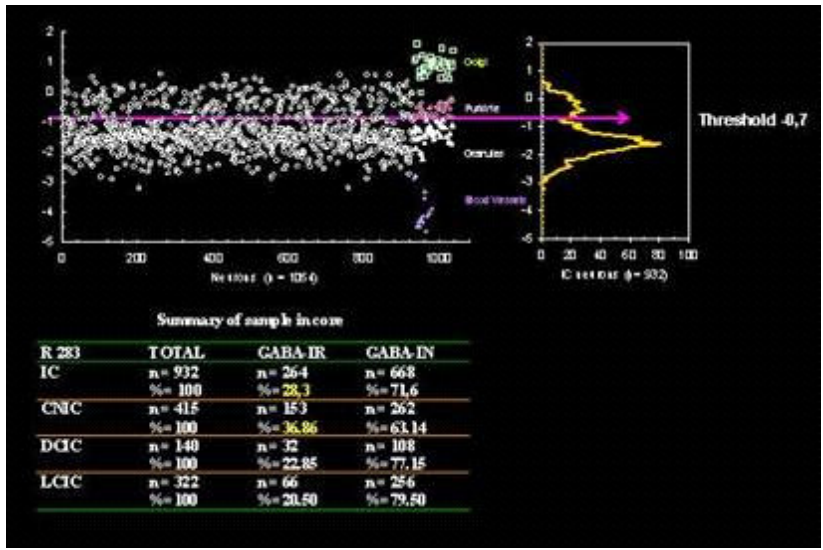

**Fig. 1.** Scatter plots showing the distribution of the grey values (normalized density) obtained for all IC neurons used in the densitometric analysis together with the control neurons (granule-, Golgi-, and Purkinje cells from the cerebellum and blood vessels). Horizontal stipple line shows the threshold calculated as the mean plus two times the S.D. of the optical density of the granule cells. The vertically oriented histogram shows a bimodal distribution of the same values seen in the scatter plot for the IC neurons. Note that the valley that separates the two peaks in the histograms approximates the threshold of OD that separates GABA-IR from GABA-IN neurons

When the study was performed in the CNIC, we found that the distribution of GABAergic neurons was different within each isofrequency band laminae and the differences across frequency regions were almost negligible (see fig 2).

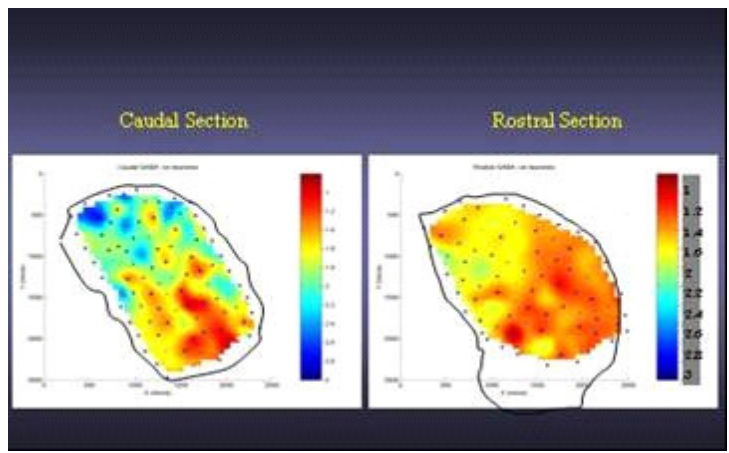

**Fig. 2.** Graphic-representation method of density gradient (GABA-IR in CNIC) codified by color. We can observe a clear topological distribution in rostral and caudal sections. More details see reference [15].

The different distributions of neurons in every tonotopic plane are correlated with a GABA gradient concentration in dorso-ventral direction. These discoveries could be interpreted as a topological relationship with laminar organization in the Central Nucleus of Inferior Colliculus. The data shows the GABAergic neuron clusters oriented almost transversal to isofrequency bands (see fig. 3).

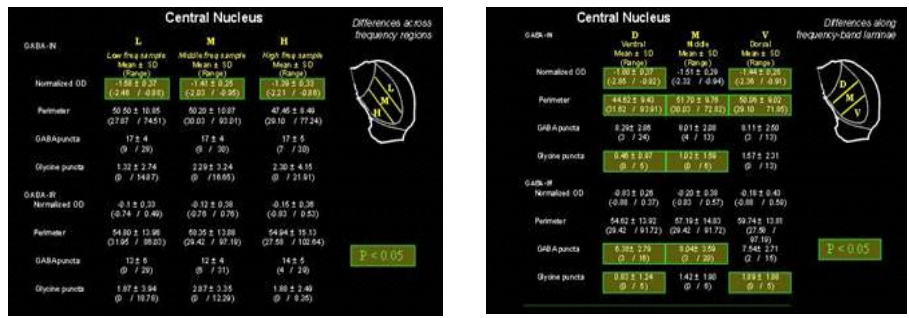

**Fig. 3.** The data shows the distribution of GABA-IR and GABA-IN neurons. Compare the difference between normalized OD, perimeter, GABA puncta and Gly puncta (in green) along and across frequency regions.

## 3. Evolvable Complex Information

"Evolvability is an organism's capacity to generate heritable phenotypic variation" [7]. Once a linkage between evolvability and robustness is established, we can see neutral mutation as the key for increased levels of phenotypic variability, which enhance the chances of innovative roads in evolution [27]. One of these innovations is the evolvable increase of complexity in a trait like information management by neural aggregations.

We begin with certain clarifications on the information notion, the central characteristic of a theoretical construct that we developed to study pre-biotic world emergence. Our construct, called the Informational Dynamic System, contains the essential capacities of autonomy, function, and information [22].

In this way, our conceptual exploration leads us to realize that the neurons manage and process - with certain levels of success - different kinds of signals or signs. And when they are processing and transmitting a message through the neuronal circuit, they incorporate some degree of 'meaning' in these actions.

Concerning information, any type of signal or sign can be a carrier of what might be information (it is the carrier of 'potential information'). We consider a signal or sign to be any matter-energy variation (sound wave, electromagnetism, concentration of a chemical compound, change in pH, etc.). The sign or signal that is a 'potential information' carrier must be in the surroundings, on the inside, or have transmitted the information to some system component.

The important thing for our idea is that the information, properly said – The Information – *has a meaning* (very basic semantic) that is created on the inside of the system. As Menat would say, "It is meaningful information." [14].

In a naturalist perspective, any signal, sign, or information must always be with respect to something else and not imposed by some outside observer. Therefore, we do not accept the existence of things like signs, signals, or information by themselves.

As with the notions of autonomy and function, we could state that the information notion is also a relational concept and, together with the other two notions, defines the Informational Dynamic Systems:

- The emergence of information (meaningful information) is made possible because of the previous existence of a matter-energy variation. What is also needed is the existence of a kind of system having the capacity of processing the matter-energy variation incident that influences it. Then, the meaningful information (or properly stated, *Information*,) *is a resultant property produced in the system* due to its processing capacity together with the matter-energy variation incident. Inside the system, a set of processes begins for the purpose of generating certain meaning to the system because of this 'potential information' (the matter-energy variation) coming to it.

- The reasons for understanding that the system will generate a 'sense or meaning' from the matter-energy variation are found in the system's intrinsic interdependence among its different processes, which produces - in one way or another - a counterbalance or a decrease/increase of the performance conditions in a process, which will have in turn certain influences over one or more processes in the system. It is in this interconnected and interdependent network of processes that we can find the constraint that makes it possible for the system to be in the far from thermodynamic equilibrium state.

- For the purposes of our work, we will use Collier's notion of cohesion [4]. Cohesion is the idea that would provide the informational dynamic system its identity, in all its transformations in time. Such cohesion is constituted by the group of relationships and interdependencies that exist among the processes; it is a relational and dynamic definition that encompasses the nature of the system organization.

- Thus, that possible information that the signal carries must have been incorporated into some process. It is in that particular process where the information might be transmitted.

- Then, the 'potential information' becomes Information (*information with meaning* for the system) since it has the capacity to produce something (an effect) in the process that incorporated it, or in some other process that is directly or indirectly connected to the initial process that incorporated it, or with something of the constrictions that are effectively maintaining the system far from thermodynamic equilibrium. The effect has a repercussion in the system, influencing its own dynamic organization.

- The effect of the information that has meaning for the system can be in the maintenance or the increase of the system cohesion. As well, the effect could produce some level of interference in the system cohesion, possibly interrupting one or more processes. It is clear that meaningful information can be caused by some signal (that carries potential information) coming from the environment like a signal that is generated in the internal dynamic of the system. In all cases, whether an effect in favor of or in contrast to cohesion, the system will develop some type of response that will be correlated to that meaningful information and the process or processes enveloped by the effect.

It is reasonable to think that through evolution the neurons are becoming those cellular entities that explore the potentialities of electromagnetic field management. In this respect, we support our studies in the results and proposals which, precisely, claim that those things known as mental phenomena are found in the endogenous electromagnetic field produced by the brain [13, 21].

One of the basic characteristics of the first ancestors of living systems, those that opened the door of pre-biotic world, was the capacity to generate "meaningful information" about their environment and about their dynamical internal milieu.

Since neurons are dynamic systems and descendents of the Informational Dynamic System in our proposal, their dynamism will carry metabolic informational-functional messages to its interior, with its environment, and with other neuronal connections. At the end, each subdivision within the isofrequency band laminae will react in different dynamic ways with the incident acoustic tone.

If a dynamic topological laminar organization relationship exists, then it is possible to think of the presence of a *temporal variability excitatory flow*.

We can state, according to the open-ended evolution that is produced, that hierarchical constraint classes will appear, keeping their original objective but on a growing scale: they will generate a search to maintain, on the whole, the conditions for the survival of a living systems, as a far from thermodynamic equilibrium system.

To maintain and to improve the recognizing registers of internal and environmental conditions and to maintain and to improve the recognizing registers of the successive changes that could arise in the internal and environmental conditions, it would involve determined kinds of constraint hierarchies.

We could state the following: it is the increase in the constraint levels that make the incremental expression of semantic levels possible in more sophisticated ways on neural integrations, whose complexity levels increase in an open-ended evolution.

## 4. Conclusions

If we agree that, in the long run of an open-ended evolution, the neurons are the cells specialized in transmitting the signals of different matter-energy variations coming from the environment by means of the digital action potentials between neurons, and if we furthermore assume that these transmissions "add" certain ways of recognizing the different matter-energy variations, then it is possible to achieve some clarifications about the emergence of mental properties: (1) we can ask what role would the inter-spike variability in neurons play and (2) what role would the different distributions of neural types in a determined sensorial information transmission play?

We propose that our findings in the CNIC, together with the central thesis of our theoretical construct, are revealing ways of developing new research on mental properties. One direction is to propose understanding how the possible constituents of the neural code are related among themselves following certain associative rules. Our proposal on this issue is an attempt to contribute to the investigations in Biosemantics [5, 8, 16, 17].

Let's focus on an ideal concept of language for a moment. Grammar is linked by rules that allow us to distinguish an adequate relationship between words in a certain language. So, in this ideal language, we would have the criteria to differentiate the construction of adequate sentences from that of non adequate sentences because we already know what the word order is for constructing a sentence in this language [9-10].

In essence, we claim that studying the ways in which sensorial information is transmitted and flows will lead us to develop certain interpretations that relate brain dynamics with mental activity.

If we had reports that the sensorial information flow is correlated with determined classes of topologies in specific portions inside the sensorial transmission signal roads, together with studies on the organizational architecture of these sensorial transmission signals roads showing us that there is a difference in the spatial-temporal topology distribution of the types of neurons, then, taking into account these experimental results and the consequences derived from our conceptual construct, we could conclude something about mental properties.

An interesting point is that, in this paper, we ourselves present something like what has just been suggested.

Indeed, the constraints of the acoustic information flow due to the topological distribution of neural populations control the ways in which the information is transmitted.

The specific distribution of neural cell types (in particular, inhibitory neurons) producing gradients of inhibition and/or excitatory signals are linked, we assume, to mental rules: *the grammar of the mind?*

Actually, this data would show some vestiges, certain indications on how this mental grammar - registered and frozen in these interrelated articulation of neural types and networks - expresses the rules in which the fires, silences and 'delays/accelerations' of the acoustic information are incorporated in the brain's endogenous electromagnetic field as basic or low-level mental representations.